\documentclass[aip,apl,reprint,amsmath,amssymb]{revtex4-2}
\usepackage{graphicx}
\usepackage{booktabs}
\usepackage[colorlinks=true]{hyperref}
\graphicspath{{Figures/}}

\usepackage{color}
\usepackage[usenames, dvipsnames]{xcolor}

\usepackage{siunitx}
\DeclareSIUnit{\ueV}{\micro\electronvolt}
\DeclareSIUnit{\um}{\micro\meter}
\DeclareSIUnit{\nm}{\nano\meter}

\begin{document}

\title{Shubnikov--de Haas Characterization of Superconductor--Semiconductor Heterostructures}

%\author{Microsoft Quantum}
\author{A. M. Zimmerman}
\email{azimmerman@microsoft.com}
\author{Saeed Fallahi}
\author{Sergei Gronin}
\author{Tyler Lindemann}
\author{Patrick Sohr}
\author{Ray Kallaher}
\author{Alejandro Alcaraz Ramirez}
\author{Georg W. Winkler}
\author{Samuel M. L. Teicher}
\author{William Cole}
\author{Sebastian Heedt}
\author{Eoin O'Farrell}
\author{Gijs de Lange}
\author{Roman Lutchyn}
\author{Michael J. Manfra}
\author{John Watson}
\affiliation{Microsoft Quantum, Redmond, Washington 98052, USA}

\date{\today}

\begin{abstract}
Hybrid superconductor-semiconductor nanostructures are a central component for research spanning condensed matter physics and quantum information processing. Continued progress relies critically on the ability to characterize, control, and optimize several intrinsic material properties including spin-orbit coupling, band offsets, and disorder in a device-relevant stack that necessarily couples the electronic states of a superconducting metal film and a semiconductor. Here we report a new method to extract fundamental material parameters utilizing simple Shubnikov-de Haas (SdH) oscillation measurements in heterostructures in which metallic electronic states are coupled to a two-dimensional electron gas (2DEG) residing in an InAs quantum well beneath an aluminum thin film. Proper analysis of the full magnetoresistance data facilitates extraction of the quantum well carrier density, spin-orbit coupling strength, and both transport and quantum scattering times. Most importantly, the extracted scattering times in the 2DEG are impacted by the metal-semiconductor coupling strength allowing us to quickly gain information on proximity-induced superconducting gap without any fabrication or mK measurements. The wealth of information that is accessed with these simple measurements positions this methodology as an important tool for hybrid materials optimization. 
\end{abstract}

\maketitle

%% ===== INTRODUCTION =====
Hybrid superconductor--semiconductor (super-semi) two-dimensional electron gas (2DEG) systems are an active area of research in a wide range of fields spanning Josephson junction arrays~\cite{bottcher2018, bottcher2023}, Andreev bound state spectroscopy \cite{haxel2023, elfeky2025}, Andreev molecules~\cite{matsuo2023}, superconducting qubits \cite{casparis2018, sagi2024, liu2025}, topological superconductivity and qubits~\cite{fornieri2019, dartiailh2021, aghaee2023, aghaee2025nature, aghaee2025arxiv}, and novel microwave circuits~\cite{phan2023,hao2024}. Progress in all these fields hinges critically on the ability to accurately characterize, control, and optimize  material properties of the hybrid system including carrier density, spin-orbit coupling (SOC) and disorder in the semiconductor quantum well (QW) and super-semi coupling in the hybrid system. While there exist many simple characterizations of 2DEGs and superconducting films in isolation such as Hall bar (HB) or Van der Pauw (VdP) measurements, material properties of the composite systems have historically relied on fabrication of nanostructures which are subsequently characterized in transport experiments in dilution refrigerators~\cite{shabani2016, kjaergaard2016, kjaergaard2017, moehle2021}. This approach leads to long feedback cycles that hamper materials optimization efforts. Additionally, fabrication of nanostructures from the starting planar systems can introduce additional sources of disorder, for instance from un-optimized fabrication processes, thereby complicating the assessment of the as-grown heterostructure.

\begin{figure*}[t]
\includegraphics[width=\textwidth]{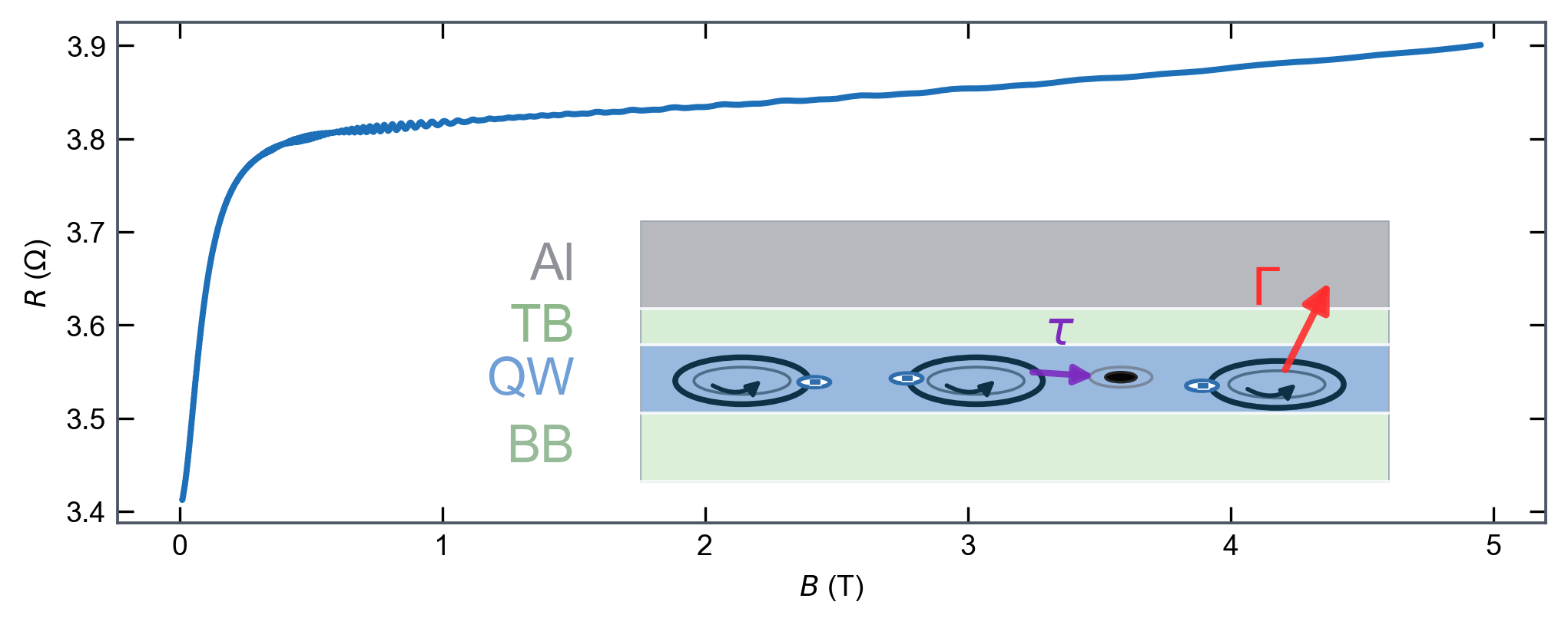}
\caption{Raw magnetoresistance $R(B)$ showing SdH oscillations on a magnetoresistance background from the combined semiconductor metal system. Data is from a sample presented in Fig. \ref{fig:gap}. Inset: schematic of the heterostructure, Al/top barrier (TB)/QW/back barrier (BB), with cyclotron orbits in the QW. Arrows indicate disorder scattering ($\tau$) and escape into the metal ($\Gamma$).}
\label{fig:raw_data}
\end{figure*}

To circumvent these challenges, we demonstrate a simple transport measurement technique relying on Shubnikov--de Haas (SdH) oscillations measured in a 2DEG under a metal film. The key insight is that the hybrid system can be treated as two parallel coupled electronic bands, much like the case of multiple occupied bands in a semiconductor. In this case, the metal can be thought of as a band with extremely high density of electrons, which results in a low resistivity, a Hall coefficient close to zero, and a lack of visible SdH oscillations. The QW has comparatively low density ($\approx 10^{12}$~cm$^{-2}$) and higher resistivity, resulting in visible SdH oscillations and a finite Hall coefficient.  The electrons in the QW undergo normal scattering from disorder in the semiconductor and metal-semiconductor interface. In addition, there is a new scattering process in which an electron from the semiconductor scatters into the metal film. The rate of this process is given by $2\Gamma/\hbar$, where $\Gamma$ is the metal-semiconductor coupling ~\cite{dassarma1985,vanheck2016}. To first order the multiple scattering processes of scattering into the metal and back to the semiconductor are suppressed~\cite{kiendl2019, lutchyn2012, potter2011}. The stack and scattering process are illustrated by the inset in Figure~\ref{fig:raw_data}.

\begin{figure*}[t]
\includegraphics[width=\textwidth]{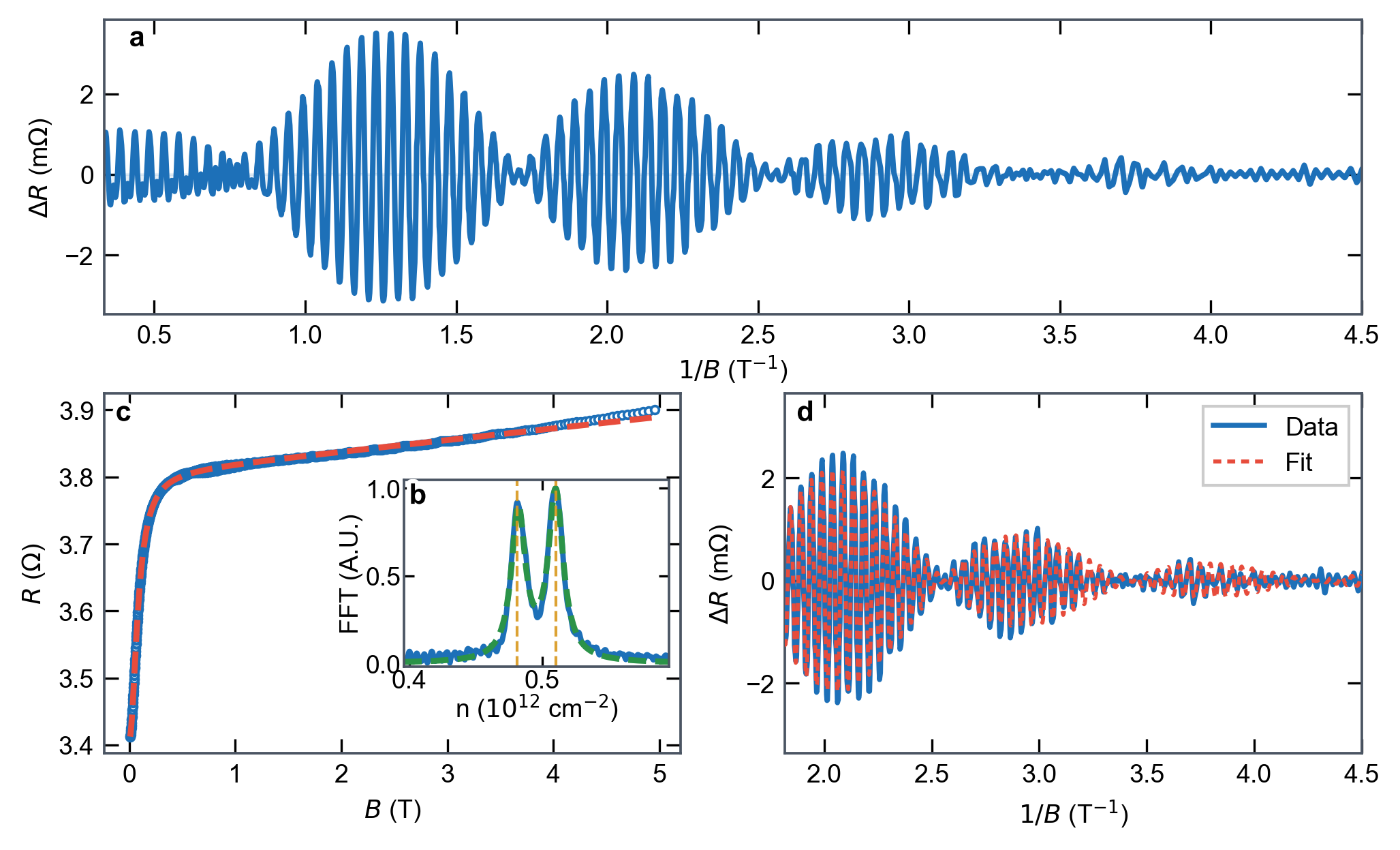}
\caption{SdH analysis steps for the data shown in Fig. \ref{fig:raw_data}. (a)~Background-subtracted oscillations $\Delta R$ vs.\ $1/B$ showing beating from spin-orbit coupling. (b)~Inset of panel~c: FFT of $\Delta R(1/B)$ for B $< 1$ T, showing two spin-split peaks and Lorentzian fit (green). (c)~Raw $R(B)$ with two-channel magnetoresistance fit (red dashed). The fit is performed over the entire data range (d)~$\Delta R$ vs.\ $1/B$ with full SdH model fit (red, dashed) used to extract $\tau_q$. Fit is performed for B $< 0.7$ T.}
\label{fig:analysis}
\end{figure*}

The scattering into the metal is a parallel scattering channel to the regular disorder scattering in the semiconductor. Because of this, the lifetimes we measure in the hybrid system are given by
\begin{equation}
\frac{1}{\tau_{t(q)}} = \frac{1}{\tau_{t(q)\text{ semi}}} + \frac{2\Gamma}{\hbar}.
\label{eq:tau_sum}
\end{equation}
where $\tau_{t(q)}$ denotes the measured transport (quantum) scattering time and $\tau_{t(q)\text{ semi}}$ are the standard lifetimes due to disorder scattering of electrons in the semiconductor. We use the effect of the metal-semiconductor coupling on the semiconductor lifetimes as a way to gain information about the induced superconducting gap in the hybrid system when the metal has transitioned to the superconducting state.

Because there is no oscillating component of the resistance in the metal, standard SdH analysis techniques can be applied to extract semiconductor QW properties; in addition, a standard model of parallel two band conduction can be fit to extract the metal and semiconductor resistivity. These measurements require no nanofabrication and are performed at temperatures above 1~K, enabling rapid feedback cycles for materials optimization.

%% ===== DEVICE STRUCTURE & METHODS =====
Our measurements are performed on a hybrid system consisting of an Al metal film deposited on a semiconductor stack with an InAs quantum well (QW) grown by molecular beam epitaxy (MBE). Unpatterned 2D films are electrically contacted by directly wire bonding to the Al film. We find that the SdH oscillations are more visible when measuring four-terminal longitudinal magnetoresistance $R(B)$ with four inline contacts instead of a traditional VdP configuration, due to larger and more uniform current in the contact region.  We measure at $T = 1.7$~K, to ensure the Al is in the normal state, over magnetic fields $B = 0$--$9$~T. A bias current of 10 to $100~\mu$A is used to improve the signal to noise ratio of the milliohm SdH oscillations on a several Ohm background. Since the metal is one or two orders of magnitude lower resistance than the QW, most of the current passes through the metal. In addition, the current density is low due to the large area of our unpatterned films, so the large current does not affect the SdH oscillations. 

\begin{figure}[t]
\includegraphics[width=0.5\textwidth]{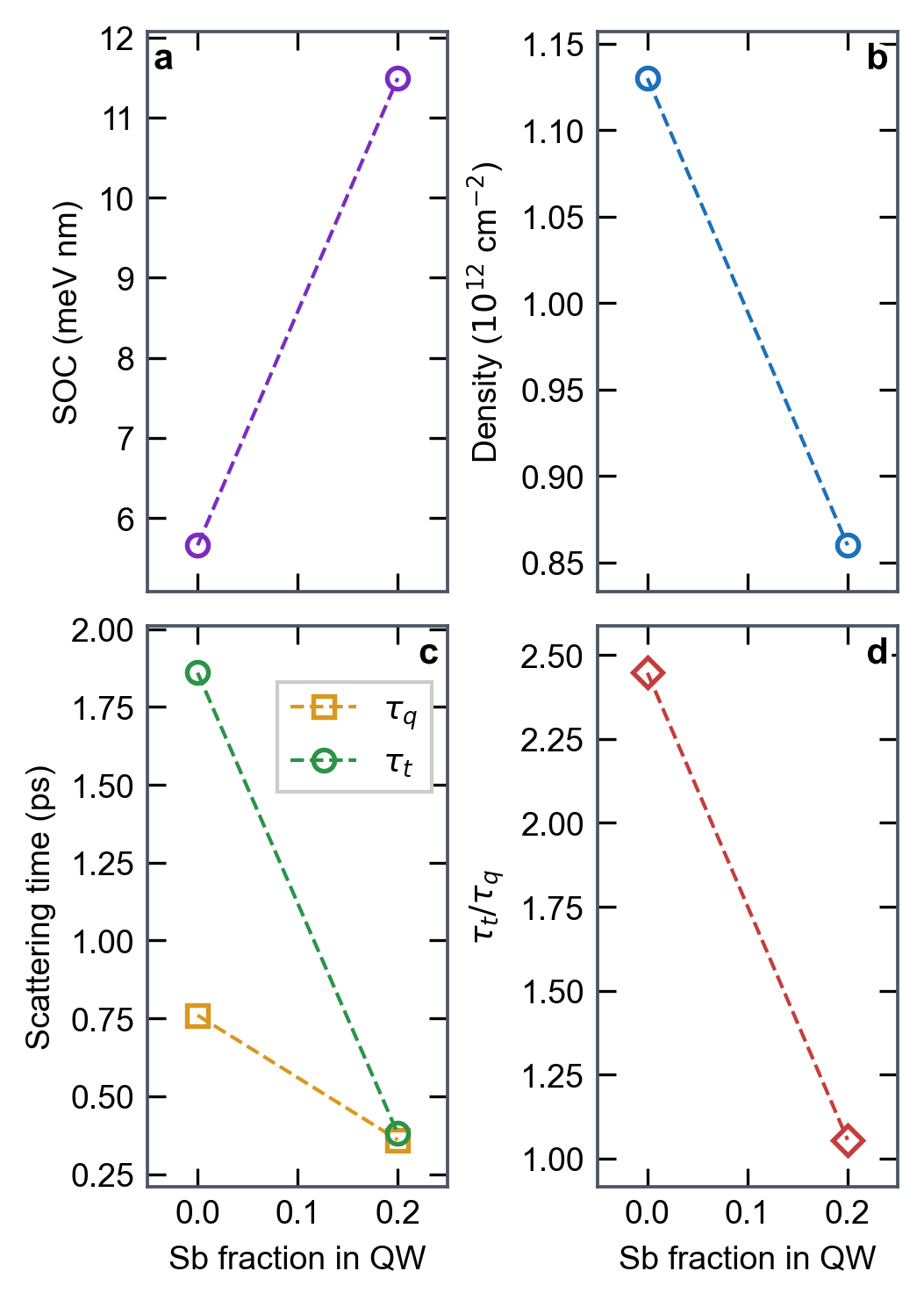}
\caption{SdH-extracted parameters vs.\ Sb fraction in the QW for InAs$_{1-x}$Sb$_x$  quantum wells. (a)~Rashba SOC $\alpha$. (b)~Carrier density $n$. (c)~Transport ($\tau_t$, circles) and quantum ($\tau_q$, squares) scattering times. (d)~Ratio $\tau_t/\tau_q$.}
\label{fig:sb_series}
\end{figure}

Figure~\ref{fig:raw_data} shows a representative $R(B)$ trace. Small SdH oscillations are visible on a large positive magnetoresistance background. Beating is visible in the oscillations due to semiconductor SOC. The background arises from the parallel conductive channels of the normal-state metal and the semiconductor QW, and can be fit with a standard model of two-band conductivity ~\cite{ashcroft1976}
\begin{equation}
\rho = \frac{\rho_S \rho_M (\rho_S + \rho_M) + (\rho_S R_M^2 +\rho_M R_S^2) B^2}{(\rho_S + \rho_M)^2 + (R_S + R_M)^2 B^2},
\label{eq:two_channel}
\end{equation}
where $\rho_{S,M}$ and $R_{S,M}$ are the resistivities and Hall coefficients of the semiconductor and metal, respectively. Due to the extremely high carrier density in a metal, we set $R_M = 0$. Then, at zero field this equation reduces to the resistivities of the two layers added in parallel and at large magnetic fields, the hybrid system resistivity approaches the metal resistivity $\rho_M$. The curvature between these two regimes depends sensitively on $\rho_S$ and $R_S$.

%% ===== ANALYSIS METHOD =====

The analysis process is illustrated in Fig.~\ref{fig:analysis}. First, the background is removed by a high-pass filter applied in the $1/B$ domain, suppressing spectral components corresponding to densities below $0.2 \times 10^{12}$~cm$^{-2}$ and yielding the oscillatory component $\Delta R(1/B)$ [Fig.~\ref{fig:analysis}(a)]. With the background removed, we can clearly see beating due to the small density differences in the Rashba spin-split subbands. Examining the amplitude envelope of the oscillations, at low field the amplitude grows as the Dingle and thermal damping factors weaken; at high field the amplitude decreases as the semiconductor Hall resistance grows, redistributing current into the lower-resistance metal channel in agreement with Eq. \ref{eq:two_channel}. At fields significantly larger than 1 T, Zeeman spin splitting and quantum Hall physics begin to onset, further reducing the oscillation amplitude.  

The FFT of $\Delta R(1/B)$ reveals two peaks corresponding to the spin-split subband densities $n_1$ and $n_2$ [Fig.~\ref{fig:analysis}(b) inset], related to the SdH frequencies ($F_i$) by $n_i = e F_i / h$~\cite{lifshitz1956}. The total density is $n = n_1 + n_2$, and the Rashba SOC parameter is~\cite{winkler2003,engels1997}
\begin{equation}
\alpha = \frac{\hbar^2 (n_h - n_l)}{m^*}\sqrt{\frac{\pi}{4\,n_l}},
\label{eq:soc}
\end{equation}
where $n_{h,l}$ are the higher and lower subband densities.

After extracting the density, the transport scattering time $\tau_t$ is extracted by fitting the raw $R(B)$ to Eq.~\eqref{eq:two_channel} [Fig.~\ref{fig:analysis}(c)]. The semiconductor Hall coefficient $R_S = 1/ne$ is fixed by the FFT density; the fit parameters are $\rho_S$, a geometric factor to convert the measured resistance to resistivity, and $\rho_M$, which we expand to be $\rho_M = \rho_{M0}+\rho_{M1}B$, adding a small magnetoresistance to account for residual field dependence at high field. The transport lifetime can be calculated from $\rho_S$ as
\begin{equation}
\tau_t = \frac{m^*}{n e^2 \rho_S}.
\label{eq:tau_t}
\end{equation}

Finally the quantum scattering time $\tau_q$ is extracted by fitting the full $\Delta R(1/B)$ waveform to the standard model for SdH oscillations [Fig.~\ref{fig:analysis}(d)]:
\begin{widetext}
\begin{equation}
\rho_S = \rho_{S_0}\!\left(1 + e^{-\pi/\omega_c \tau_q}\,\frac{Am^*T}{\sinh(Am^*T)}\sum_{i=1}^{2}\cos\!\left(\frac{2\pi^2\hbar n_i}{eB} + \phi_i\right)\right),
\label{eq:sdh_model}
\end{equation}
\end{widetext}
where $A = 2\pi^2 k_B/(e\hbar B)$ and $\omega_c = eB/m^*$~\cite{lifshitz1956,ihn2010,ando1982,coleridge1989}. A full waveform fit is necessary because the beating from the two spin-split subbands precludes a simple Dingle envelope analysis. $\tau_t$ is primarily sensitive to large-angle scattering with each scattering event at angle $\theta$ weighted by $(1-\cos\theta)$ while $\tau_q$ captures all scattering events equally. Their ratio $\tau_t/\tau_q$ therefore gives us information about the dominant scattering mechanisms: $\tau_t/\tau_q \gg 1$ indicates small-angle (long range) scattering, while $\tau_t/\tau_q \approx 1$ indicates isotropic short-range scattering~\cite{harrang1985,dassarma1985,hwang2008, dassarma2014}.

\begin{figure}[b]
\includegraphics[width=0.5\textwidth]{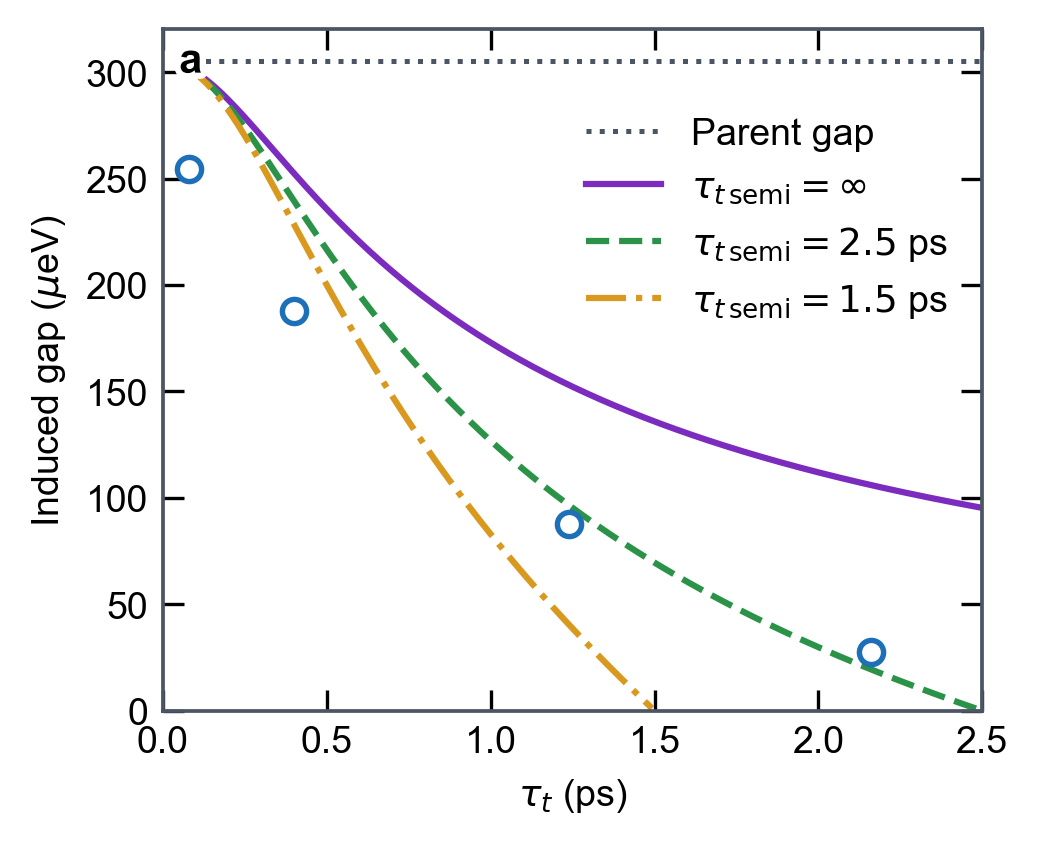}
\caption{Induced superconducting gap $\Delta_{\text{ind}}$ vs.\ transport scattering time $\tau_t$. Data points (circles) are independent QPC and SdH measurements on the same stacks. Curves show the model [Eqs.~\eqref{eq:tau_sum}--\eqref{eq:gamma}] for $\tau_{t \text{ semi}} = \infty$ (clean limit), 1.5, and 2.5~ps with $\Delta_0 = 305~\mu$eV.}
\label{fig:gap}
\end{figure}
%% ===== RESULTS: Sb  =====

To experimentally test the ability of this technique to resolve simultaneous changes in 2DEG density, SOC, and scattering, we tested the effect of adding Sb to the QW, comparing an InAs QW to an InAs$_{0.8}$Sb$_{0.2}$ QW,  Figure~\ref{fig:sb_series}. For simple comparison, to demonstrate the technique, we use an effective mass $m^* = 0.03\,m_e$ to analyze both data sets, though there will be a small change in $m^*$ due to the change in QW composition.

The changes in the measured quantities agree very well with our expectations. Spin-orbit coupling [Fig.~\ref{fig:sb_series}(a)] increases from $5.66$~meV$\cdot$nm to $11.5$~meV$\cdot$nm, as the added Sb enhances the Rashba coefficient~\cite{winkler2016}. The carrier density [Fig.~\ref{fig:sb_series}(b)] decreases modestly from $1.13$ to $0.86 \times 10^{12}$~cm$^{-2}$, consistent with the higher conduction band edge of InAs$_{0.8}$Sb$_{0.2}$ reducing charge transfer into the QW. Both $\tau_t$ and $\tau_q$ decrease  with the added Sb [Fig.~\ref{fig:sb_series}(c)], consistent with the introduction of short-range alloy scattering. In the pure InAs QW, $\tau_t = 1.86$~ps and $\tau_q = 0.76$~ps, giving $\tau_t/\tau_q = 2.45$, indicative that long-range (remote impurity) scattering is dominant. In the InAs$_{0.8}$Sb$_{0.2}$ QW, $\tau_t$ drops to $0.38$~ps while $\tau_q$ decreases to $0.36$~ps, yielding $\tau_t/\tau_q = 1.05$ [Fig.~\ref{fig:sb_series}(d)]. This convergence toward unity is characteristic of isotropic short-range alloy scattering becoming the dominant mechanism. 

We note that the lifetimes extracted here are different than those that can be measured in traditional HB or VdP measurements on a bare semiconductor at the same density. In addition to the contribution to the lifetime from the metal-semiconductor coupling, the addition of the metal on top of the semiconductor stack provides a different screening environment for charge disorder and removes contributions from the bulk dielectric, semiconductor-dielectric interface, or native oxides that are present in HBs. Like the other properties measured with this technique, the lifetimes extracted are better representations of the combined metal-semiconductor system than those extracted from HBs or VdPs.

%% ===== RESULTS: INDUCED GAP  =====
To further test the usefulness of this technique, we examined a second series of samples with intentionally varied metal-semiconductor coupling.  We do this by varying the Al concentration in the QW barriers, keeping other compositions fixed. 

As highlighted in Eq.~\ref{eq:tau_sum}, the lifetimes we extract in our measurements are affected by the metal-semiconductor coupling, $\Gamma$.
The same coupling determines the proximity-induced superconducting gap in our samples. The relationship is given by~\cite{vanheck2016}
\begin{equation}
\Gamma = \Delta_{\text{ind}}\sqrt{\frac{\Delta_0 + \Delta_{\text{ind}}}{\Delta_0 - \Delta_{\text{ind}}}}.
\label{eq:gamma}
\end{equation}

To establish this relationship empirically, we fabricate quantum point contacts (QPCs) on the boundary of a half-plane of Al (an S-QPC-N geometry~\cite{kjaergaard2016}) and perform bias spectroscopy measurements to extract $\Delta_{\text{ind}}$. SdH measurements are performed on separate samples cleaved from the same wafers using the SdH method described above. For this comparison, we use the transport lifetime, $\tau_t$, which is longer than $\tau_q$ and therefore more constrained by the metal-semi coupling.

Figure~\ref{fig:gap} plots the induced gap $\Delta_{\text{ind}}$ against the SdH transport lifetime $\tau_t$ for this series of stacks. The curves show the prediction of induced gap based on measured $\tau_t$ and the metal-semiconductor coupling, calculated for different disorder scattering times, $\tau_{t \text{ semi}}$, [Eqs.~\eqref{eq:tau_sum}--\eqref{eq:gamma}]. We use a parent gap $\Delta_0 = 305~\mu$eV determined from QPC measurements tunneling into the Al half-plane and consistent with previous measurements in similar devices \cite{aghaee2023}. The data are consistent with this model and fall along the curve for $\tau_{t \text{ semi}} = 2.5$~ps, indicating a comparable disorder scattering time across these structurally similar stacks and allowing us to estimate the disorder scattering in the semiconductor. More importantly, $\tau_t$ provides an upper limit for the induced gap which can be further constrained by estimates of disorder.  This is a powerful tool for materials development allowing quick feedback for induced gap without requiring separate fab and measurement of QPC devices.

%% ===== CONCLUSION =====
In summary, we have reported SdH oscillations in a 2DEG underneath an Al metal film, providing direct access to the carrier density, spin-orbit coupling, and scattering times of the semiconductor QW in the device-relevant heterostructures. The extracted scattering times are constrained by the metal-semiconductor coupling, allowing us to provide fast feedback on the proximity-induced superconducting gap. These measurements require no nanofabrication, are performed at temperatures above 1~K, and probe the 2DEG without removing the metal. By enabling rapid extraction of key semiconductor parameters from as-grown planar heterostructures, this technique can significantly accelerate materials optimization across the broader super-semi research community, from Josephson junction devices to topological qubits and beyond.

\begin{acknowledgments}
We thank the Microsoft Quantum team for support and discussions.
\end{acknowledgments}

\section*{Author Contributions}

\textbf{A. M. Zimmerman}: Conceptualization (lead), formal analysis (lead), methodology (equal), investigation (lead), software (lead), writing/original draft preparation (equal), writing -- review \& editing (equal).
\textbf{Saeed Fallahi}: Investigation (supporting), Writing -- review \& editing (supporting).
\textbf{Sergei Gronin}: Resources (equal), writing -- review \& editing (supporting).
\textbf{Tyler Lindemann}: Resources (equal), writing -- review \& editing (supporting).
\textbf{Patrick Sohr}: Resources (equal), writing -- review \& editing (supporting).
\textbf{Ray Kallaher}: Resources (equal), writing -- review \& editing (supporting).
\textbf{Alejandro Alcaraz Ramirez}: Resources (equal), writing -- review \& editing (supporting).
\textbf{Georg W. Winkler}: Methodology (supporting), writing -- review \& editing (supporting).
\textbf{Samuel M. L. Teicher}: Methodology (equal), writing -- review \& editing (supporting).
\textbf{William Cole}: Methodology (equal), writing -- review \& editing (supporting).
\textbf{Sebastian Heedt}: Methodology (equal), writing -- review \& editing (supporting).
\textbf{Eoin O'Farrell}: Methodology (equal), writing -- review \& editing (supporting).
\textbf{Gijs de Lange}: Supervision (equal), Methodology (equal), writing -- review \& editing (supporting).
\textbf{Roman Lutchyn}: Supervision (equal), methodology (equal), writing -- review \& editing (supporting).
\textbf{Michael J. Manfra}: Supervision (equal), writing -- review \& editing (equal).
\textbf{John Watson}: Supervision (equal), writing/original draft preparation (equal), writing -- review \& editing (equal).

\bibliographystyle{aipnum4-2}
\bibliography{references}

\end{document}